\documentclass{ws-ijmpe}

\usepackage{graphicx}

\begin{document}

\markboth{P.~Van~Isacker}
{Partial seniority conservation}

\catchline{}{}{}{}{}

\title{PARTIAL CONSERVATION OF SENIORITY IN NUCLEI}

\author{P.~VAN~ISACKER}

\address{Grand Acc\'el\'erateur National d'Ions Lourds, CEA/DSM--CNRS/IN2P3\\
B.P.~55027, F-14076 Caen Cedex 5, France\\
isacker@ganil.fr}

\maketitle

\begin{history}
\received{(received date)}
\revised{(revised date)}
\end{history}

\begin{abstract}
A review is given of the relation 
between pairing, quasi-spin algebras and seniority.
The former two concepts are closely connected,
the relation being that the quasi-spin formalism
allows an efficient solution of the pairing problem.
Seniority is of much wider applicability than either pairing or quasi-spin
and applications can even be extended
to encompass the notion of partial symmetry.
A recent application of partial conservation of seniority
and its importance for nuclear isomers is discussed.
\end{abstract}

\section{Pairing, seniority and quasi-spin algebras}
\label{s_intro}
The pairing interaction is a reasonable first-order approximation
to the strong force between identical nucleons.
For nucleons in a single-$j$ shell,
as will be assumed throughout in this contribution,
pairing is defined by the matrix elements
\begin{equation}
\nu_J\equiv
\langle j^2;JM_J|\hat V|j^2;JM_J\rangle=
-g(2j+1)\delta_{J0},
\label{e_pairj1me}
\end{equation}
where $j$ is the total (orbital+spin) angular momentum of a single nucleon
(hence $j$ is half-odd-integer),
$J$ results from the coupling of two $j$s
and $M_J$ is the projection of $J$ on the $z$ axis.
Furthermore, $g$ is the strength of the interaction
which is attractive in nuclei ($g>0$).

The pairing interaction was introduced by Racah
for the classification of $n$ electrons in an atom\cite{Racah43}.
He was able to derive a closed formula
for the interaction energy among the $n$ electrons
and to prove that any eigenstate of the pairing interaction
is characterized by a `seniority number' $\upsilon$
which corresponds to the number of electrons
that are not in pairs coupled to orbital angular momentum $L=0$.
Racah's original definition of seniority
made use of coefficients of fractional parentage.
He later noted that simplifications arose
through the use of group theory\cite{Racah49}.
Seniority turned out to be a label
associated with the (unitary) symplectic algebra ${\rm Sp}(2j+1)$
in the classification
\begin{equation}
\begin{array}{ccccc}
{\rm U}(2j+1)&\supset&{\rm Sp}(2j+1)&\supset&{\rm SU}(2)\\
\downarrow&&\downarrow&&\downarrow\\[0mm]
[1^n]&&[1^\upsilon]&&J
\end{array}.
\label{e_clasj1}
\end{equation}
Since the nucleons are identical,
all states of the $j^n$ configuration belong to
the totally antisymmetric irreducible representation (IR) $[1^n]$ of ${\rm U}(2j+1)$.
The IRs of ${\rm Sp}(2j+1)$
therefore must also be totally antisymmetric of the type $[1^\upsilon]$
with allowed values of seniority $\upsilon=n,n-2,\dots,1$ or 0.

In the definition~(\ref{e_clasj1}),
seniority appears as a label associated with the algebra ${\rm Sp}(2j+1)$.
This has the drawback that, depending on $j$,
the algebra can be quite large.
Matters become even more complicated
when the fermions are non-identical
and are assigned an intrinsic label $t$ with $2t+1$ substates,
such as is the case with nucleons which have isospin $t={\frac 1 2}$.
The total number of single-particle states is then $\Omega\equiv(2j+1)(2t+1)$
and the classification~(\ref{e_clasj1}) is generalized to\cite{Racah52,Flowers52}
\begin{equation}
\begin{array}{ccccccccccc}
{\rm U}(\Omega)&\supset&\Bigl({\rm U}(2j+1)&\supset&{\rm Sp}(2j+1)&\supset&{\rm SU}_J(2)\Bigr)
&\otimes&\Bigl({\rm U}(2t+1)&\supset&{\rm SU}_T(2)\Bigr)\\
\downarrow&&\downarrow&&\downarrow&&\downarrow&&\downarrow&&\downarrow\\[0mm]
[1^n]&&[h]&&[\sigma]&&J&&[\tilde h]&&T
\end{array},
\label{e_clasjt}
\end{equation}
where $[h]$, $[\tilde h]$ and $[\sigma]$ are Young tableaux
associated with ${\rm U}(2j+1)$,  ${\rm U}(2t+1)$ and ${\rm Sp}(2j+1)$, respectively,
$J$ is the total angular momentum and $T$ the total isospin
(or, in general, the total coupling of all $t$s).
The classification~(\ref{e_clasjt}) assumes a symmetry ${\rm SU}_J(2)\otimes{\rm SU}_T(2)$,
that is, an interaction which conserves angular momentum $J$ and isospin $T$.
To ensure overall antisymmetry under ${\rm U}(\Omega)$,
the Young tableaux of ${\rm U}(2j+1)$ and ${\rm U}(2t+1)$ must be {\em conjugate},
that is, $[\tilde h]$ is obtained from $[h]$
by interchanging rows and columns.
This conjugate relationship can be expressed as follows:
\begin{equation}
[h]=[\overbrace{2t+1,2t+1,\dots,2t+1}^{\tilde h_{2t+1}},
\overbrace{2t,2t,\dots,2t}^{\tilde h_{2t}-\tilde h_{2t+1}},\dots,
\overbrace{1,1,\dots,1}^{\tilde h_1-\tilde h_2}],
\label{e_h1}
\end{equation}
and, since $[h]$ contains at most $2j+1$ labels,
this proves that overall antisymmetry requires $\tilde h_1\leq2j+1$.
The converse relation reads
\begin{equation}
[\tilde h]=[\overbrace{2j+1,2j+1,\dots,2j+1}^{h_{2j+1}},
\overbrace{2j,2j,\dots,2j}^{h_{2j}-h_{2j+1}},\dots,
\overbrace{1,1,\dots,1}^{h_1-h_2}],
\label{e_h2}
\end{equation}
and shows that $h_1\leq2t+1$.
The requirement of overall antisymmetry
sets limits on the allowed tableaux $[h]$ or $[\tilde h]$.
Assuming for the sake of definiteness that $t\leq j$,
the Young tableau of ${\rm U}(2j+1)$
can be specified by $2t+1$ labels $\tilde h_i$
instead of the $2j+1$ normally needed.
As a consequence of the form~(\ref{e_h1}) of $[h]$,
all non-zero labels in $[\sigma]$ must be $2t+1$, $2t$, \dots or 1,
and the Young tableau of ${\rm Sp}(2j+1)$ must therefore be of the form
\begin{equation}
[\sigma]=[\overbrace{2t+1,2t+1,\dots,2t+1}^{\sigma_{2t+1}},
\overbrace{2t,2t,\dots,2t}^{\sigma_{2t}},\dots,
\overbrace{1,1,\dots,1}^{\sigma_1}].
\end{equation}
The IR of ${\rm Sp}(2j+1)$ is thus characterized by $2t+1$ labels at most.
For identical nucleons, $t=0$, there is one label,
namely seniority $\upsilon$.
For non-identical nucleons, $t={\frac 1 2}$, there are two such labels,
introduced by Flowers\cite{Flowers52},
which correspond to seniority $\upsilon$ and to reduced isospin $t_\upsilon$.
The former has the same interpretation as in the like-nucleon case
while the latter corresponds to the isospin of the nucleons
which are not in pairs coupled to $J=0$.

It is clear that one quickly runs into formidable group-theoretical reduction problems
if one follows the formulation associated with the classification~(\ref{e_clasjt}).
Fortunately, an alternative and simpler definition of seniority can be given
in terms of algebras that do not change with $j$.
The idea was simultaneously and independently proposed
by Kerman\cite{Kerman61} for $t=0$ ({\it i.e.}, for identical particles)
and by Helmers\cite{Helmers61} for general $t$.
The starting point is the introduction of the operators
\begin{eqnarray}
\hat S^j_+(m_t,m'_t)&=&
\sqrt{2j+1}\,
(a_{j,tm_t}^\dag\times a_{j,tm'_t}^\dag)^{(0)}_0\equiv
\hat S^j_+(m'_t,m_t),
\nonumber\\
\hat S^j_-(m_t,m'_t)&=&
\left(\hat S^j_+(m_t,m'_t)\right)^\dag\equiv
\hat S^j_-(m'_t,m_t),
\nonumber\\
\hat S^j_z(m_t,m'_t)&=&
\sqrt{2j+1}\,
(a_{j,tm_t}^\dag\times\tilde a_{j,tm'_t})^{(0)}_0-
{\frac 1 2}(2j+1)\delta_{m_tm'_t}.
\label{e_gen}
\end{eqnarray}
The particle creation operators $a_{jm_j,tm_t}^\dag$
are assigned an angular momentum $j$ with projection $m_j$
and an intrinsic quantum number $t$ with projection $m_t$.
To ensure the correct transformation properties under rotations,
the operators $\tilde a_{jm_j,tm_t}\equiv(-)^{j+m_j+t+m_t}a_{j-m_j,t-m_t}$
are introduced.
Furthermore, the coupling, indicated in~(\ref{e_gen}) by $\times$,
refers to angular momentum
and no coupling in $t$ is carried out.
It can be shown\cite{Helmers61} that the operators~(\ref{e_gen})
close under commutation
and generate the (unitary) symplectic algebra ${\rm Sp}(4t+2)$.
Furthermore, these operators are `symplectic invariants'
in the sense that they commute with all the generators
of the symplectic algebra ${\rm Sp}(2j+1)$ introduced in~(\ref{e_clasjt}).
Because of this property,
the IRs of ${\rm Sp}(2j+1)$
[contained in the antisymmetric IR of ${\rm U}(\Omega)$]
are in one-to-one correspondence with those of  ${\rm Sp}(4t+2)$.
The two IRs are {\em complementary}, that is,
one is obtained from the other
by constructing the complement of the Young tableau
in a rectangle with $j+{\frac 1 2}$ rows and $2t+1$ columns
and reflecting this complement
along the lower-left-to-upper-right diagonal
of the rectangle\cite{Helmers61}.
This complementary relationship
between the $[\sigma]\equiv[\sigma_1,\sigma_2,\dots,\sigma_{j+{\scriptstyle\frac 1 2}}]$
labels of ${\rm Sp}(2j+1)$
and the $[\bar\sigma]$ labels of ${\rm Sp}(4t+2)$
can be expressed as follows:
\begin{equation}
[\bar\sigma]=[\overbrace{j+{\textstyle\frac 1 2},j+{\textstyle\frac 1 2},\dots,j+{\textstyle\frac 1 2}}^{2t+1-\sigma_1},
\overbrace{j-{\textstyle\frac 1 2},j-{\textstyle\frac 1 2},\dots,j-{\textstyle\frac 1 2}}^{\sigma_1-\sigma_2},\dots,
\overbrace{1,1,\dots,1}^{\sigma_{j-{\scriptstyle\frac 1 2}}-\sigma_{j+{\scriptstyle\frac 1 2}}}].
\label{e_sigma}
\end{equation}
For $t=0$ (identical fermions) the operators~(\ref{e_gen})
generate the algebra Sp(2) which is isomorphic to SU(2)
and which therefore has been named `quasi-spin' algebra\cite{Kerman61}.
The seniority label $\upsilon$ in the classification~(\ref{e_clasj1})
goes over into the Sp(2) label $(2j-2\upsilon+1)/2$
which corresponds to $(2j-2\upsilon+1)/4$ in SU(2).
For $t={\frac 1 2}$ (neutrons and protons)
one finds the `quasi-spin' algebra ${\rm Sp}(4)\sim{\rm SO}(5)$,
extensively studied by Hecht\cite{Hecht65}.
The seniority and reduced isospin $[2^{\upsilon/2-t_\upsilon},1^{2t_\upsilon}]$
in the classification~(\ref{e_clasjt})
go over in the Sp(4) labels $[(2j-\upsilon+2t_\upsilon+1)/2,(2j-\upsilon-2t_\upsilon+1)/2]$
which correspond to $[(2j-\upsilon+1)/2,t_\upsilon]$ in SO(5).

The connection with pairing is established
by constructing the operator
\begin{equation}
\hat Q=
{\frac 1 4}\sum_{m_tm'_t}
\hat S^j_+(m_t,m'_t)\hat S^j_-(m_t,m'_t),
\end{equation}
which is related to the Casimir operator of ${\rm Sp}(4t+2)$ through
\begin{equation}
\hat Q=
{\frac 1 4}n(\Omega-n+2)-
\frac{1}{16}\Omega(4t+2j+5)+
{\frac 1 4}\hat C_2[{\rm Sp}(4t+2)],
\end{equation}
where use is made of the complementarity relation
between ${\rm Sp}(2j+1)$ and ${\rm Sp}(4t+2)$.
Its eigenvalues are found from those of $\hat C_2[{\rm Sp}(4t+2)]$ which are
$\sum_i\bar\sigma_i(\bar\sigma_i+4t+4-2i)$.
For $t=0$, $\hat Q$ coincides with the pairing interaction, $\hat V_{\rm P}=\hat Q$,
and the following eigenvalue expression for the pairing interaction results:
\begin{equation}
E_{nj}(\upsilon)={\frac 1 4}(n-\upsilon)(2j-n-\upsilon+3).
\end{equation}
For $t={\frac 1 2}$, the relation between $\hat Q$ and $\hat V_{\rm P}$ becomes
\begin{equation}
\hat V_{\rm P}=
\hat Q-
{\frac 1 2}\hat C_2[{\rm SU}_T(2)]-
{\frac 1 8}n(4j-n),
\end{equation}
and leads to the eigenvalue expression for the isospin-invariant pairing interaction:
\begin{equation}
E_{nj}(T,\upsilon,t_\upsilon)=
{\frac 1 8}(n-\upsilon)(4j-n-\upsilon+8)
-{\frac 1 2}[T(T+1)-t_\upsilon(t_\upsilon+1)].
\end{equation}

This succinct and simultaneous derivation of the eigenvalues
in the cases of identical nucleons and of neutrons and protons,
illustrates the power of the quasi-spin formulation of the pairing problem.

\section{Conservation of seniority}
\label{s_consen}
While the preceding analysis shows
that the eigenstates of a pairing interaction
carry the quantum numbers associated with seniority,
it does not answer the question
what are the necessary and sufficient conditions
for a general interaction to conserve seniority.
This problem has been worked out in its full generality
for identical particles ({\it i.e.}, $t=0$).
In that case, a rotationally invariant two-body interaction $\hat V$
is completely specified by the matrix elements $\nu_J$ in Eq.~(\ref{e_pairj1me})
with $J=0,2,\dots,2j-1$.
The necessary and sufficient conditions
for the conservation of seniority
can then be written as\cite{Talmi93,Rowe01,Rosensteel03}
\begin{equation}
\sum_{J=2}^{2j-1}(2J+1)\sqrt{2I+1}
\left(\frac{\delta_{JI}}{\sqrt{(2J+1)(2I+1)}}
+2\Bigl\{\begin{array}{ccc}j&j&J\\ j&j&I\end{array}\Bigr\}
-\frac{4}{(4j^2-1)}
\right)\nu_J=0,
\label{e_condi}
\end{equation}
with $I=2,4,\dots,2j-1$,
and where the symbol between curly brackets is a Racah coefficient.
No condition on the matrix elements $\nu_J$
is obtained for $j=3/2$, 5/2 and 7/2,
one condition for $j=9/2$, 11/2 and 13/2, and so on.
Clearly, the conditions~(\ref{e_condi}) are much weaker
than the requirement that the interaction be of pairing character
but still many of the results of the quasi-spin formalism remain valid.

\section{Partial conservation of seniority}
\label{s_partial}
It was shown recently\cite{Isacker08}
that the seniority quantum number
occasionally can be conserved for some but not all eigenstates.
One may characterize this as a {\em partial} conservation of seniority,
akin to the notion of partial dynamical symmetry\cite{Alhassid92}.
There are a number of  `trivial' examples
of partial conservation of seniority.
For example, if the total angular momentum $J$ is odd,
a four-particle state cannot be of seniority $v=0$ or $v=2$
and must necessarily have seniority $v=4$.
Also, for $J>2j$ the four-particle state must be of seniority $v=4$.
These trivial cases are not of concern here.
The discussion of partial conservation of seniority
so far has been confined to the case of identical fermions
which is assumed in this and the following section.

A non-trivial example of partial conservation of seniority
was discovered by Escuderos and Zamick\cite{Escuderos06}
who pointed out that four identical fermions with $j=9/2$
display one $J=4$ and one $J=6$ state 
both of which have seniority $v=4$
for an {\em arbitrary} interaction.
This is an example of a partial symmetry,
where seniority is broken for most but not for all states.
An analysis of the four-particle system shows\cite{Isacker08}
that the two eigenstates can be written as
\begin{eqnarray}
|(9/2)^4,v=4,J=4\rangle&=&
\sqrt{\frac{25500}{25591}}|(9/2)^4[22]4\rangle-
\sqrt{\frac{91}{25591}}|(9/2)^4[24]4\rangle,
\nonumber\\
|(9/2)^4,v=4,J=6\rangle&=&
\sqrt{\frac{27132}{27257}}|(9/2)^4[24]6\rangle+
\sqrt{\frac{125}{27257}}|(9/2)^4[26]6\rangle.
\label{e_wave}
\end{eqnarray}
Furthermore, the two states have the eigenenergies
\begin{eqnarray}
E[(9/2)^4,v=4,J=4]&=&
\frac{68}{33}\nu_2+\nu_4+\frac{13}{15}\nu_6+\frac{114}{55}\nu_8,
\nonumber\\
E[(9/2)^4,v=4,J=6]&=&
\frac{19}{11}\nu_2+\frac{12}{13}\nu_4+\nu_6+\frac{336}{143}\nu_8.
\label{e_energ}
\end{eqnarray}
The wave functions~(\ref{e_wave}) are identical to those of Eq.~(8) of Ref.\cite{Isacker08}
but written here in a Gram-Schmidt basis.
For the definition of this basis one starts, for $J=4$,
from the non-orthogonal set $|(9/2)^4[II']4\rangle$
with $[II']=[04]$, [22] and [24].
(The labels $I$ and $I'$ refer to the angular momenta of the two pairs
as they appear in a two-fermion-pair basis\cite{Isacker08}.)
The second and third states
obtained after the Gram-Schmidt orthonormalization
are orthogonal to $|(9/2)^4[04]4\rangle$,
and hence by definition have seniority $v=4$.
A similar argument is valid for $J=6$
where the non-orthogonal set $|(9/2)^4[II']6\rangle$
has $[II']=[06]$, [24] and [26].
Consequently, the states~(\ref{e_wave}) have seniority $v=4$.

In summary, the wave functions of the two states~(\ref{e_wave}) are pre-determined
and their energies are linear combinations
of the interaction matrix elements $\nu_J$
with coefficients that are rational non-negative numbers.
These results are valid for an {\em arbitrary} interaction among $j=9/2$ fermions.
According to  Talmi's definition\cite{Talmi10},
the states are therefore solvable,
independent of whether the interaction conserves seniority or not.
Although a mathematical derivation can be given
of the necessary conditions for the existence of such solvable states,
a simple, intuitive reason for it is still lacking.
Some progress towards this goal has been made\cite{Zamick08},
and  a partial understanding
of the coefficients entering the energy expressions~(\ref{e_energ})
has been achieved.

The property of partial conservation of seniority
may appear rather academic
because it only occurs four identical fermions in a single shell with $j=9/2$.
Nevertheless, as argued in the next section,
this peculiar exception is crucial for the existence of seniority isomers in nuclei.

\section{Seniority isomers in semi-magic nuclei}
\label{s_isomer}
In nuclei, states with low seniority generally occur at low energy.
For example, the ground state of an even--even semi-magic nucleus
has approximately $\upsilon\approx0$
(all nucleons in pairs coupled to $J=0$)
while its lowest levels with angular momenta $J=2,4,6,\dots$
usually have $\upsilon\approx2$ (containing one `broken' pair with $J\neq0$).
Seniority isomerism is expected to occur in semi-magic nuclei
because electric quadrupole (E2) transitions between $\upsilon=2$ states
are hindered, especially when the valence shell is close to half-filled.
This result is a consequence of the fact that
the matrix elements of the quadrupole operator
between states with seniority $\upsilon=2$ vanish at mid shell\cite{Talmi93}.

Examples of seniority isomers have been found
in the $N=50$ isotones
with protons dominantly confined to the $1g_{9/2}$ shell.
In particular, the $J^\pi=8^+$ levels
in $^{94}$Ru ($Z=44$) and $^{96}$Pd ($Z=46$)
have half-lives of 71 and 2.2~$\mu$s, respectively,
resulting from a combination of slow E2 decay
and a small energy difference with the $J^\pi=6^+$ levels below it.
A review is given by Grawe {\it et al.}\cite{Grawe97}.
On the basis of similar arguments
one would expect the same phenomenon to occur
in the neutron-rich nickel ($Z=28$) isotopes $^{72}$Ni and $^{74}$Ni
with neutrons dominantly confined to the $1g_{9/2}$ shell
but this does not seem to be the case\cite{Sawicka03}.

The seniority classification is a good approximation
for $n$ identical nucleons in a $j=9/2$ shell.
This is a trivial statement
if only one state exists
for a given particle number $n$ and angular momentum $J$,
in which case seniority must be exact.
If two states with different seniority and the same $J$ occur,
it can be shown that for any reasonable nuclear interaction $\hat V$
the off-diagonal matrix element is small
compared to the states' energy difference.
The proof that seniority mixing is negligible
for the $J^\pi=4^+$ and $6^+$ states
of a $(9/2)^4$ system is more subtle
and is crucial for the existence of seniority isomers.
There are {\em three} states for each of these angular momenta,
two of which, with $\upsilon=2$ and $\upsilon=4$, are close in energy
and could possibly strongly mix.
However, the $\upsilon=4$ members of these closely-spaced doublets
are the $J^\pi=4^+$ and $6^+$ states discussed in Sect.~\ref{s_partial}
which {\em have exact seniority $\upsilon=4$ for any interaction}.
As a result, breaking of seniority only arises through the mixing
between the $\upsilon=2$ and the higher-lying $\upsilon=4$ states
and, by the same argument as above,
this mixing is found to be small.

\begin{figure}
\includegraphics[width=12.5cm]{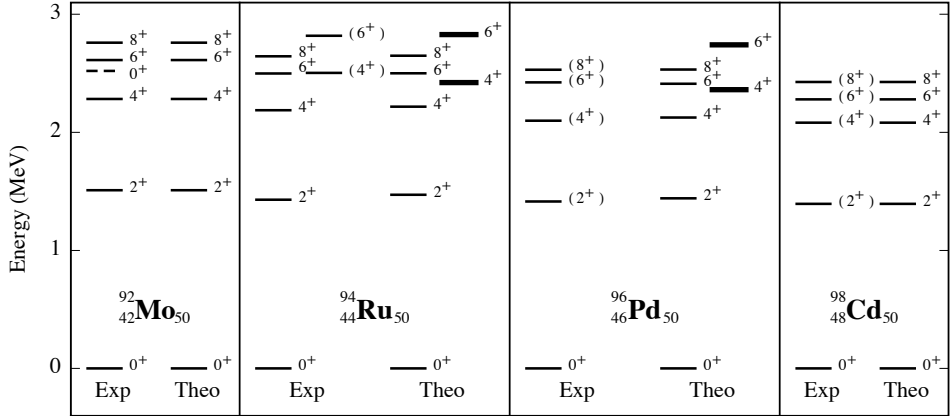}
\caption{The low-energy spectra of the $N=50$ isotones
$^{92}$Mo, $^{94}$Ru, $^{96}$Pd and $^{98}$Cd.
The left-hand parts of each panel show the observed levels
while the right-hand parts
contain the results of a $(1g_{9/2})^n$ shell-model calculation
with linearly varying two-body matrix elements.
The $J^\pi=4^+$ and $6^+$ levels with exact $\upsilon=4$
are shown with thick lines;
the dashed line corresponds to an intruder level.}
\label{f_n50}
\end{figure}
\begin{figure}
\includegraphics[width=12.5cm]{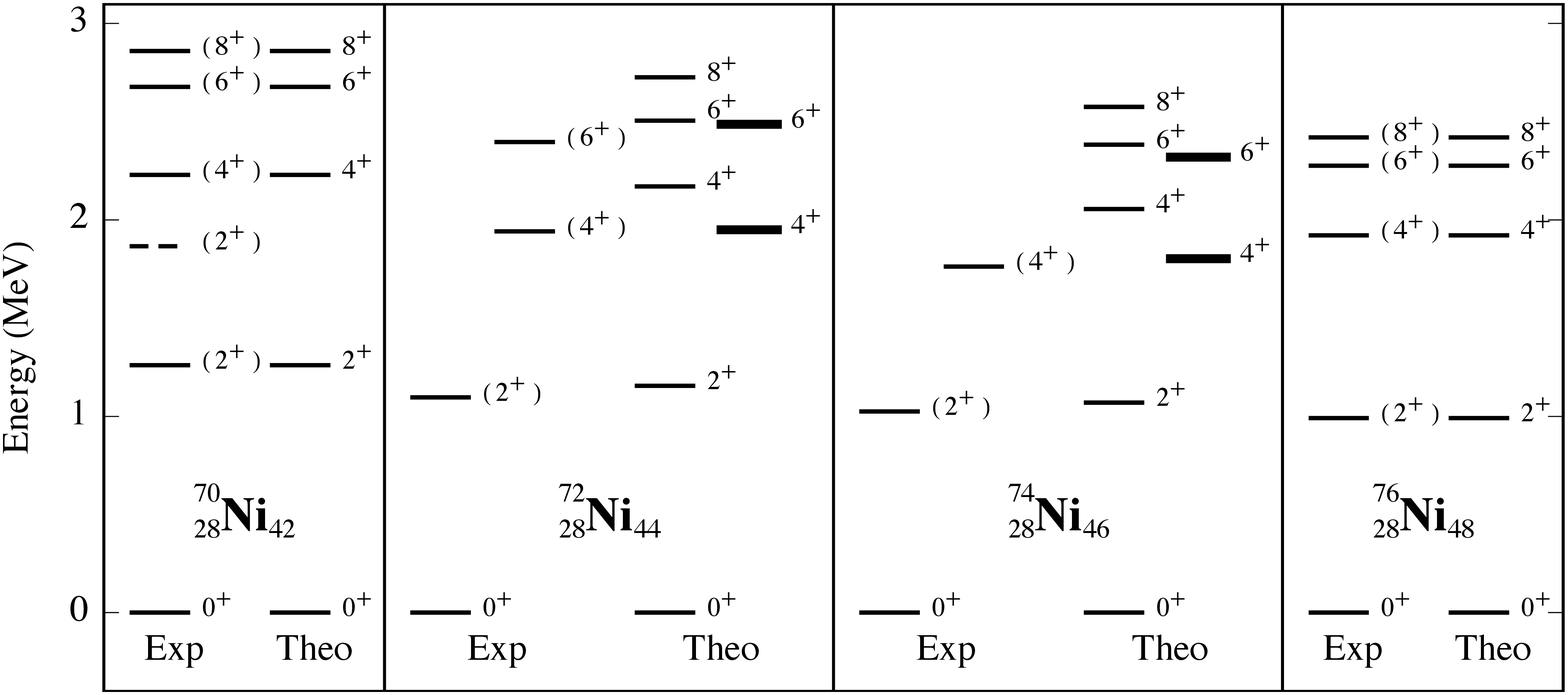}
\caption{The low-energy spectra of the nickel isotopes $^{70-76}$Ni.
See caption of Fig.~\ref{f_n50}.}
\label{f_ni}
\end{figure}

In Fig.~\ref{f_n50} the results of a calculation for the $N=50$ even--even isotones
are compared to the observed spectra\cite{NNDC}.
The two-particle and the two-hole spectra, $^{92}$Mo and $^{98}$Cd,
determine the two-body matrix elements $\nu_J$.
Different procedures can be followed to fix these matrix elements,
all leading to similar results,
and those shown in Fig.~\ref{f_n50}
are obtained by varying $\nu_J$ linearly between $^{92}$Mo and $^{98}$Cd.
The description of the two intermediate isotopes, $^{94}$Ru and $^{96}$Pd,
should then be rather accurate, albeit very phenomenological.
In $^{94}$Ru $(4^+_2)$ and $(6^+_2)$ levels
are observed at energies of 2503 and 2818~keV, respectively\cite{Mills07};
these are possible candidates for the states with exact $\upsilon=4$ (thick lines)
which are calculated at 2422 and 2828~keV.
In Fig.~\ref{f_ni} are shown
the observed spectra of the nickel isotopes\cite{NNDC},
compared to a calculation similar to the one performed for the $N=50$ isotones.

There is a striking difference
between the calculated four-particle and four-hole spectra
of the $N=50$ isotones and those of the nickel isotopes:
the $\upsilon=4$, $J^\pi=4^+$ and $6^+$ states
are yrast in $^{72}$Ni and $^{74}$Ni
but not in $^{94}$Ru and $^{96}$Pd.
This drastically changes the E2-decay pattern.
The `typical' decay with small $B$(E2) values between $\upsilon\approx2$ states
is characteristic of the seniority classification
in nuclei near mid-shell ($n\approx j+{\frac 1 2}$)
and is at the basis of the explanation
of seniority isomers\cite{Grawe97}.
This situation applies to $^{94}$Ru and $^{96}$Pd
where the $\upsilon\approx2$ states are yrast.
On the other hand,
the E2 decay pattern, as it is calculated in $^{72}$Ni and $^{74}$Ni
where the yrast $J^\pi=4^+$ and $6^+$ levels have $\upsilon=4$,
leads to fast E2 decay in the yrast band.
As a consequence, the $J^\pi=8^+$ and $6^+$ levels
are unlikely to be isomeric in this case.

Given that approximate analytic expressions
are available for the energies of all relevant states,
conditions for the existence of isomers
can easily be established.
This shows that their existence
is predominantly governed by the quadrupole pairing matrix element $\nu_2$
of the nucleon--nucleon interaction\cite{Isackerun}.

\section{Summary}
\label{s_conc}
In this contribution the relation
between pairing, seniority and quasi-spin algebras
was reviewed.
It was argued that seniority is of broader applicability
than pairing or quasi-spin.
In a few special cases,
the conservation of seniority turns out to be partial,
and this has important implications
for the existence of seniority isomers in nuclei.

\end{document}